\documentclass[aps,prl,twocolumn,superscriptaddress]{revtex4-2}
\usepackage{graphicx}
\usepackage{amsmath}
\usepackage[urlcolor=blue]{hyperref}
\hypersetup{colorlinks=true, linkcolor=blue, citecolor=blue}

\begin{document}


\title{Probing the electronic topological transitions of WTe$_2$ under pressure using ultrafast spectroscopy}

\author{Kai Zhang}
\affiliation{Key Laboratory of Materials Physics, Institute of Solid State Physics, HFIPS, Chinese Academy of Sciences, Hefei 230031, China}
\affiliation{GBA Branch of Aerospace Information Research Institute, Chinese Academy of Sciences, Guangzhou 510700, China}
\author{Fuhai Su}\email{fhsu@issp.ac.cn}
\affiliation{Key Laboratory of Materials Physics, Institute of Solid State Physics, HFIPS, Chinese Academy of Sciences, Hefei 230031, China}
\author{Dayong Liu}\email{dyliu@ntu.edu.cn}
\affiliation{Department of Physics, School of Sciences, Nantong University, Nantong 226019, China}
\author{Wenjun Wang}
\affiliation{Anhui Key Laboratory of Condensed Matter Physics at Extreme Conditions, High Magnetic Field Laboratory, Hefei Institutes of Physical Sciences, Chinese Academy of Sciences, Hefei, 230031, Anhui, China}
\author{Yongsheng Zhang}
\affiliation{Key Laboratory of Materials Physics, Institute of Solid State Physics, HFIPS, Chinese Academy of Sciences, Hefei 230031, China}
\author{Zhi Zeng}
\affiliation{Key Laboratory of Materials Physics, Institute of Solid State Physics, HFIPS, Chinese Academy of Sciences, Hefei 230031, China}
\author{Zhe Qu}\email{zhequ@hmfl.ac.cn}
\affiliation{Anhui Key Laboratory of Condensed Matter Physics at Extreme Conditions, High Magnetic Field Laboratory, Hefei Institutes of Physical Sciences, Chinese Academy of Sciences, Hefei, 230031, Anhui, China}
\author{Alexander F. Goncharov}\email{agoncharov@carnegiescience.edu}
\affiliation{Earth and Planets Laboratory, Carnegie Institution for Science, Washington, D.C. 20015, USA}

\date{\today}

\begin{abstract}
We investigate the nonequilibrium photocarrier dynamics of WTe$_2$ under pressure using the optical pump-probe spectroscopy. The pressure dependences of the electronic relaxation manifest anomalous changes around 0.8, 3.5, and 6 GPa, indicating the abruptions in the electron-phonon interactions. In addition, the coherent phonon oscillations originating from shear mode suddenly disappears above 3.5 GPa, which marks the onset of Td-1T’ structural phase transition. Supported by the theoretical calculation, we unveil the electronic topological transitions (ETTs), especially an emergence of a new type-II Weyl point for Td-WTe$_2$ under pressure. Our work demonstrates a novel route to probe the ETTs under pressure.
\end{abstract}


\maketitle


Dirac and Weyl semimetals possess linear energy dispersions and massless fermions, as well as the a high mobility, ultralow scattering rate,\cite{Bansz,Liang,Ali} and thus are instructive in understanding basic concepts in high-energy physics.\cite{Weyl} The Dirac points commonly appear in highly symmetrical materials, however, are able to evolve into two Weyl points with opposite chirality once the inversion symmetry or time-reversal symmetry is broken.\cite{Yan} Moreover, the Dirac and Weyl semimetals can be divided into two types (types I and II), depending on the tilting of the band structure around the neutral point.\cite{Yan} Weyl semimetals with band-touching points can host extraordinary physical phenomena, such as negative magnetoresistance and topological Fermi arcs.\cite{Huang,Yao,Xu} These aforementioned versatile properties make the Weyl semimetals attractive both in the fundamental research and device applications.\cite{Ali,Qian,Jiang-1,Tokur} Pressure provides a clean route to continuously tune the interactions between atoms, which can give rise to the modification of the electronic structure without concomitant lattice abruptions, featuring electronic topological transitions (ETT) or emergence of new Dirac/Weyl points in materials.\cite{Cheng,Xiao,Xiang} Generally, highly symmetrical Dirac/Weyl points can be directly identified using angular resolved photoelectron spectroscopy (ARPES) by mapping the electronic energy bands. ARPES, however, cannot be implemented under extreme conditions, such as high pressure. Other experimental techniques, such as the Shubnikov-de Haas effect, electrical resistance, and Raman/infrared spectroscopy, also are used to probe the emerging Dirac/Weyl points under pressure.\cite{Xiang,Jiang-2,Zhang-2,Zhang-3} These measurements are sensitive to the electron structures around band extremes or the Fermi level, which are always subjected to the defect states and disturbances from the neighboring electron or/and hole pockets, as shown in Fig. 1(a). Hence, new diagnostic methods accessing the information of electrons away from the Fermi level or extreme points of energy bands are highly desirable to understand the ETTs under pressure.

For materials systems, including semiconductors and semimetals, the ultrafast photoexcitation using femtosecond (fs) or ps laser pulse is able to establish a transient nonequilibrium electronic state, where the photocarriers are distributed over the large k space (Fig. 1 (a)). The subsequent relaxation dynamics strongly depends on the details of the electronic structures.\cite{Gierz,Othon} Such photo-induced electronic redistribution normally results in a higher temperature (T$_e$) than lattice (T$_l$), which can launch the electron-phonon (e-ph) interactions with the participation of not only the low-frequency acoustic phonons but also the high-frequency optical phonons.\cite{Othon} The ultrafast spectroscopy can be used to track the time evolution of the intraband and interband electronic transitions, which allows us to access the energy band dispersion at high-k region and disentangles the complex interactions among different quasiparticles,\cite{Gierz,Othon,Blanc} beyond what is attainable with conventional steady-state optics or transport measurements. Therefore, the ultrafast spectroscopy in combination with a diamond anvil cell (DAC), applied to the studies of nonequilibrium quasiparticles and electrons cooling for strongly correlated systems and low-dimensional materials,\cite{Zhang-1,Zhang-4} may open a new avenue to decipher the ETTs and/or the emerging Dirac/Weyl points under pressure. In recent years, diverse ultrafast spectroscopies have been employed to investigate the topological semiconductors or semimetals under ambient conditions, but only rarely have been extended to high-pressure conditions.\cite{Zhang-4} WTe$_2$, as a representative type-II Weyl semimetal,\cite{Li,Soluy} adopts the orthorhombic Td phase with distorted triangular lattices of tungsten and tellurium atoms, showing strong anisotropy at ambient condition.\cite{Xia,Zhou,Jana,Song,Frenz,Jha,Chen,Chen-1,Wei} With applying the external pressure, the Td phase is suppressed and gradually transformed to the monoclinic 1T’ structure phase, which is accompanied by the emergence of the superconductivity and the elimination of the topological Weyl fermions due to the introduction of the inversion center.\cite{Xia,Zhou,Lu,Pan,Kang,Li-1} In addition, a Lifshitz transition also occurs at around 3 GPa, introducing a series of exotic phenomena, such as the drastic weakening of the extremely large magnetoresistance, the pronounced changes of the optical conductivity spectrum, etc.\cite{Pan,Cai,Krott} Therefore, WTe$_2$ can serve as an ideal prototypical testbed for the investigations of pressure-induced ETT.

In this work, we employed the fs optical pump-probe spectroscopy (OPPS) combined with the DAC technique to investigate the photocarrier dynamics of WTe$_2$ under hydrostatic pressure up to 21.3 GPa. We find that the coherent phonon (CP) excitation arising from the interlayer shear vibration is suddenly suppressed above 3.5 GPa. Moreover, the fast time constant ($\tau$$_{fast}$) shows the discontinuity at approximately 0.8, 3.5, and 6 GPa, respectively. By combining the theoretical analysis, we reveal the ETTs under pressure, especially the emergence of the new type-II Weyl points around the Fermi level above 3.5 GPa.

For the experiment, the high-quality WTe$_2$ single crystals were grown through self-flux method.\cite{Wang,Zhu,Ali-1} In addition, we repeated the measurements using other samples, including the WTe$_2$ crystals purchased from the SixCarbon Technology and HQ Graphene, and obtained the same results. The hydrostatic pressure condition was obtained using a DAC with the culet of 300 $\mu$m, in which we applied silicon oil as the pressure-transmitting medium. All of the samples were exfoliated mechanically to obtain a fresh and flat surface before loading into the sample chamber. The pressures were gauged by the photoluminescence of the ruby balls. To access the photocarrier dynamics of WTe$_2$ under pressure, we conducted the fs-OPPS with reflectivity geometry in combination with the DAC. Briefly, the fs pulses with 1.55 eV photon energies, delivered from an fs oscillator, were used as the pump and probe sources for the fs-OPPS. The pump and probe beams were created using a beamsplitter and a controllable time delay was introduced for the probe beam. As shown in Fig. 1(b), the pump and probe beams were spatially separated; they were focused on the sample inside the DAC using an objective lens with spot diameters of 10 and 4 $\mu$m, respectively. In the measurements, the orthogonal polarizations of the pump and probe beams were maintained to efficiently suppress the background scattering signal. We rotated the DAC with the sample along the DAC axis to examine the angular dependencies of the signal.

\subsection{Pressure dependences of photocarrier dynamics}
\begin{figure*}[htbp]
\includegraphics[width=0.9\textwidth]{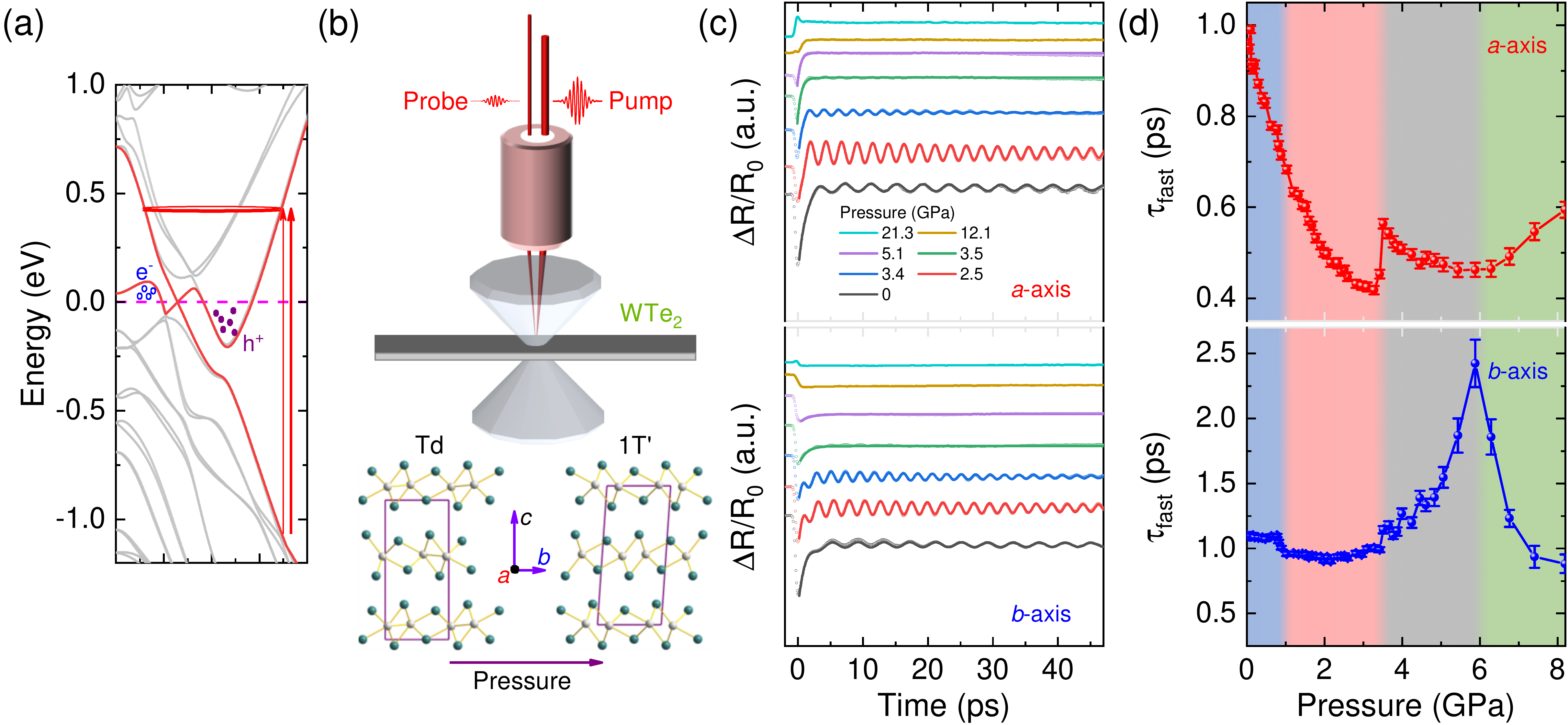}
\caption{(a) Typical hot carrier excitation and subsequent decay channels for the Dirac/Weyl materials. The thin and thick lines represent the pump and probe processes, respectively. The ring indicates the hot carriers with the Fermi-Dirac distribution. (b) Block schemes of the fs-OPPS setups. The bottom images show the lattice structures of the WTe$_{2}$ with Td and 1T' phases. (c) Normalized transient differential reflectivity $\Delta$R/R$_{0}$ of WTe$_{2}$ with the pump polarization along the a- and b-axis of the sample at different pressures. All curves have been displaced vertically for clarity. (d) Pressure dependence of the fast decay time $\tau$$_{fast}$ with the pump beam polarizations along a- and b-axes, respectively.}
\end{figure*}

The differential reflectivity (DR) transients, $\Delta$R/R$_0$, ($\Delta$R=R-R$_0$) as a function of pump-probe delay under selected pressures are presented in Fig. 1(c), where the R or R$_0$ is the reflectivity with and without the pump, respectively. More pressure-dependent data are provided in Fig. S4 in the Supplemental Material. In the measurement [Fig. 1(c)], the polarization of the probe light was set as parallel in either an a direction (upper panel) or b direction (lower panel). We determined the crystalline orientation of the Td-WTe$_2$ crystal based on the angle-resolved Raman spectroscopy and fs-OPPS (see Figs. S1 and S3 in the Supplemental Material). Hereafter, these two probe-polarization geometries are referred to as parallel and perpendicular configurations. At ambient conditions, the DR signal features an initial instantaneous dropping, which is followed by a relaxation process composed with a fast rising and a subsequent slow decay superimposed with periodic oscillations. The periodic oscillation stems from the excitation of the CP corresponding to the A$_1$ optical phonon mode, with a frequency of 0.25 THz.\cite{Dai,He} The incoherent electronic response contains the fast and slow relaxation stages; the fast component is attributed to the energy dissipation of the photocarrier through optical phonon scattering, while the slow component should be governed by the interactions between the electrons and acoustic phonons\cite{Othon} or the phonon-assisted e-h recombination between the electron and hole pockets,\cite{Homes} exhibiting a long-lived offset within our time-scanning range.

As revealed by Figs. 1(c) and S4, the photocarrier dynamics in WTe$_2$ shows obvious in-plane anisotropy due to the directional one-dimensional tungsten-chain structure, which is akin to the observations in black phosphorous.\cite{Ge} Clearly, the DR transients strongly depend on the external pressure. In addition, the two common pressure effects for both parallel and perpendicular polarization configurations can be directly identified from the Figs. 1(c) and S4. First, in both polarization configurations, the amplitude of the fast component with negative sign in the incoherent DR signal pronouncedly declines upon compressing, and almost disappears as the pressure is beyond 6 GPa. Instead, with further elevating pressure, the fast relaxation component over subpicosecond timescale resurge in the manner of rapid rising with positive sign above 15.5 GPa. Second, the CP oscillations suddenly vanish as the pressure is beyond 3.5 GPa. Previous works based on the high-pressure Raman and XRD experiments have corroborated that the adjacent Te-W-Te layers undergo an opposite sliding, leading to a structural transition from the orthorhombic Td-WTe$_2$ to monoclinic 1T’-WTe$_2$.\cite{Xia,Zhou,Lu} However, no consensus was reached as for the turning pressure of the Td-1T’ phase transition pressure. The reported transition points are scattered in the pressure ranges from 4.5 to 9.6 GPa.\cite{Xia,Zhou,Lu,Li-1,Krott} The suppression of the CP excitation of the shear mode was proposed to serve as an indicator for Td-1T’ phase transition in MoTe$_2$,\cite{Zhang} although it was not accessed yet under high pressure. Therefore, we suggest that the disappearance of the CP oscillations at 3.5 GPa marks the onset of the Td-1T' transition.

To gain the quantitative insights into the pressure dependence of the photocarrier dynamics, we fit the DR curve in terms of a damped harmonic oscillator function combined with a single-exponential function for each pressure, as follows:
\begin{equation}
    \begin{aligned}
  \frac{\Delta R}{R_0} = A_f exp (- \frac{t}{\tau_f}) + A_{CP} exp (- \frac{t}{\tau_{CP}}) cos (2 \pi f t + \varphi) + B
   \end{aligned}
\end{equation}
where A$_{CP}$ is the CP amplitude, $\tau$$_{CP}$ is the lifetime of CP, f is the CP frequency, and $\varphi$ is the initial phase of CP oscillations. A$_f$ and $\tau$$_f$ denote the amplitude and time constant of the fast decay component, respectively. According to the two-temperature model,\cite{Dai,Allen} the e-ph coupling constant, $\lambda$, is roughly proportional to the energy relaxation rate 1/$\tau$$_f$. Thereby, first of all, we pay special attention to the $\tau$$_f$ to assess the pressure evolution of the e-ph interactions. The extracted time constant for the fast decay process ($\tau$$_f$), for parallel and perpendicular configurations, versus pressure is presented in Fig. 1(d). As the probe light is along the a direction, the $\tau$$_f$ monotonously decreases with increasing pressure, while shows an abrupt discontinuous change of approximately 37$\%$ in magnitude around 3.5 GPa. With further compression, the $\tau$$_f$ again reduces with pressure up to 6 GPa, followed by a gradual rising at higher pressures. Interestingly, the pressure dependence of $\tau$$_f$ obtained in the perpendicular configuration exhibits a different pressure behavior. The $\tau$$_f$ essentially remains constant below 3.5 GPa, with an exception of a recognizable sudden drop at around 0.8 GPa. An abrupt rising of the $\tau$$_f$ around 3.5 GPa is also discernible although the value is far smaller than that in the parallel configuration. Under higher pressure, the time constant pronouncedly increases up to 6 GPa, and subsequently dramatically shortens.

Overall, we clearly distinguished the three discontinuities around 0.8, 3.5, and 6 GPa for the pressure dependence of $\tau$$_f$, emphasizing the pressure-induced abruptions in the e-ph interactions. To further reveal if there are additional lattice structure changes except the Td-1T’ phase transition, we carried out the high-pressure Raman scattering measurement with short pressure step. As exemplified in Fig. S2 in the Supplemental Material, the pressure behaviors of Raman modes do not manifest any discontinuity but a subtle uptrend on the slop of the frequency at around 3.5 GPa in our studied pressure range, which is in line with earlier reports.\cite{Xia,Zhou,Lu} In addition, we can see that the CP experiences an initial increase in the amplitude before it completely disappears as shown in Fig. S4. The totally suppression of the CP signal occurs in a sharp pressure range as 0.5 GPa, which is in striking contrast with the reported gradual evolution from Td to 1T’ phase.\cite{Xia,Zhou,Li-1} Thus, we infer that the abrupt changes in electronic structure, such as the Lifshitz transition should be taken into account to explain the anomalies of the pressure dependences of the photocarrier dynamics.

\subsection{DFT calculations for electronic structures of WTe$_2$ under pressure}
\begin{figure*}[htbp]
\includegraphics[width=0.9\textwidth]{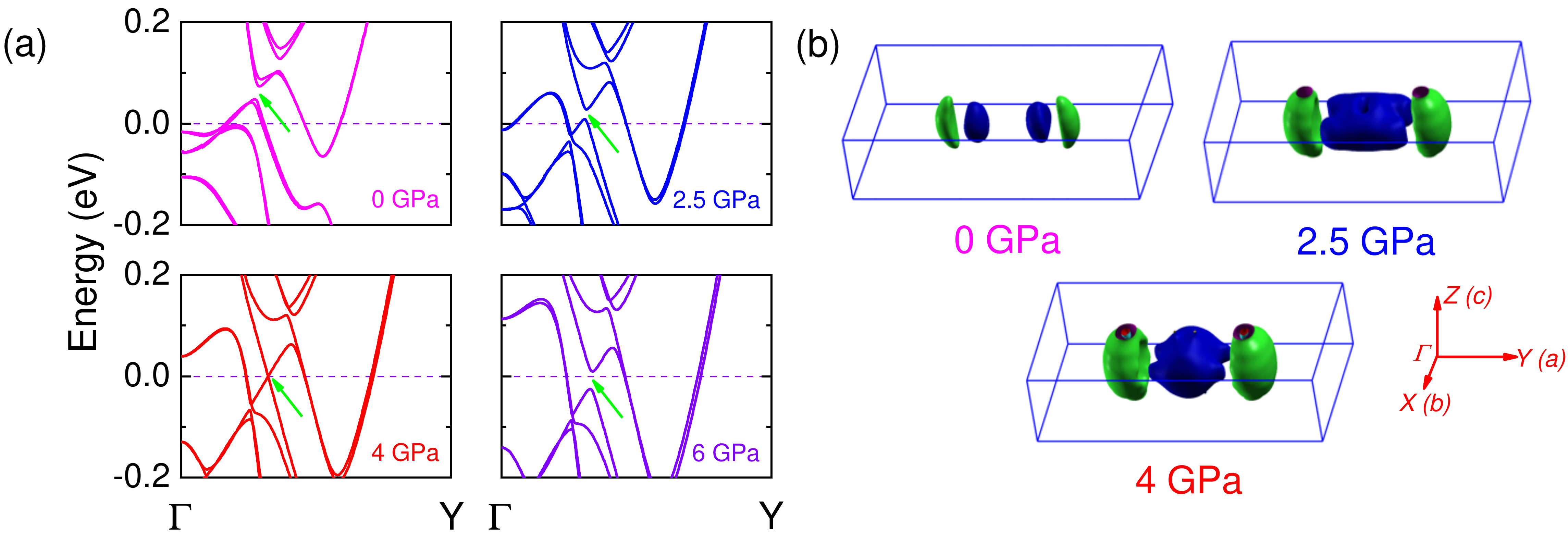}
\caption{(a) Pressure evolution of the band structure along $\Gamma$-Y direction. Bands associated with the Weyl-like point is pointed out by arrows. (b) Fermi pockets at 0, 2.5, and 4 GPa.}
\end{figure*}

To unveil the pressure effects of the electronic structures for Td-WTe$_2$, we performed the DFT calculations, (see the details in the Supplemental Material). The calculated energy bands under pressure are shown in Fig. 2(a). Accordingly, the cross sections of the Fermi surface around the Fermi level at 0, 2.5, and 4 GPa were extracted, shown in Fig. 2(b). One can find that there exist sharp changes around Fermi level (-0.1 eV$\sim$0.1 eV) along $\Gamma$-Y with increasing pressure. At the pressure of 0 GPa, the type-II Weyl points, derived from the tiny band crossing above the Fermi level at around 0.05 eV, can be recognized, which is consistent with a previous report.\cite{Soluy} As the pressure increases up to about 2.5 GPa, both the electron and hole pockets gradually grow in size along with increasing pressure, and eventually the two electron pockets merge into one, as shown in Fig. 2(b). Such a topological change of the Fermi surface has been observed by the magneto-transport methods,\cite{Cai} demonstrating a Lifshitz transition. The growth of the pocket volumes implies an augment of the density of electronic states at k-space around the energy-band extreme points, which can effectively render more available phase space at low-energy range for hot carrier distribution and thus shorten the fast decay time constant.\cite{Zhang-1,Zhang-4} Remarkably, the electronic density of states (DOS) along the $\Gamma$-X direction in k-space (along the b-axis in real space) surges due to the merging of the electron pockets, and thus can lead to a sudden drop of the $\tau$$_f$ as the probe polarization is along the b-axis. Note that the $\tau$$_f$ is not sensitive to the adjustment of the pump polarization, as shown in Fig S3(a) in the Supplemental Material. This observation seems plausible because the initial orientating momenta of the photocarriers, endowed by the polarized pump photon upon photoexcitation, would be rapidly randomized over a few tens of femtoseconds (shorter than the pulse duration of the fs laser used in our experiment) through carrier-carrier scatterings.\cite{Malic}

Further increasing the pressure up to 4 GPa, our calculations reveal that a new Weyl point with linear band dispersion emerges at around the Fermi level, accompanying with a gap closing. At further compression, this new Weyl point vanishes, and a gap reopens but the Fermi surface topology retains when the pressure significantly increases up to about 6 GPa. This new Weyl point is quite different from that at low pressure (0$\sim$1 GPa). The initial eight Weyl points originate from the double degeneracies due to the spin-orbit coupling,\cite{Soluy} while the high-pressure Weyl points result from the point contact between the conduction and valence bands modulated by the external pressure. This result implies possibly distinct transport or ultrafast dynamical behaviors at different pressures. Since the e-ph interaction for the electronic band with a linear energy dispersion is commonly expected to be weakened due to the large Fermi velocity and small effective mass,\cite{Si} we propose that the formation of Weyl point can facilitate the decoupling between electron and phonon, resulting in the sudden increase of $\tau$$_f$ around 3.5 GPa. The anisotropy of the new-born Weyl points in k-space should be responsible for the polarization sensitivity of the $\tau$$_f$. Especially, the $\tau$$_f$ exhibits a reversed pressure behavior above 3.5 GPa for parallel and perpendicular configurations (Fig. 2(d)). Such an effect is consistent with the shift of the new Weyl points from the highly symmetrical $\Gamma$-Y direction (a-axis) to the $\Gamma$-X direction (b-axis) as unveiled by the calculation. We should mention that our calculations do not reveal sudden change in the electronic structure around 6 GPa for WTe$_2$ with Td phase, and cannot account for the turning point of 6 GPa of the pressure dependence of the $\tau$$_f$. We propose that there are two possible mechanisms underlying the anomaly. (1) The gradual Td-1T’ phase transition may lead to the hybrid phases with Td and 1T’ structural WTe$_2$ in the 4-6 GPa pressure range. This anomaly in photocarrier dynamics can be trigged when the 1T’ phase dominates with pressure. (2) Additional ETT likely takes place for 1T’ phase at around 6 GPa, resulting in the abruption of e-ph interaction. Overall, the anomalous transitions of $\tau$$_f$ with pressure should be correlated with the electronic structure transition, the first anomaly around 0.8 GPa is associated with the Lifshitz transition of the electronic bands, and the sudden changes above 3.5 GPa has a close correspondence with the emergence of the new Weyl point at the Fermi level.

\subsection{Pressure dependence of the coherent phonon}
\begin{figure*}[htbp]
\includegraphics[width=0.85\textwidth]{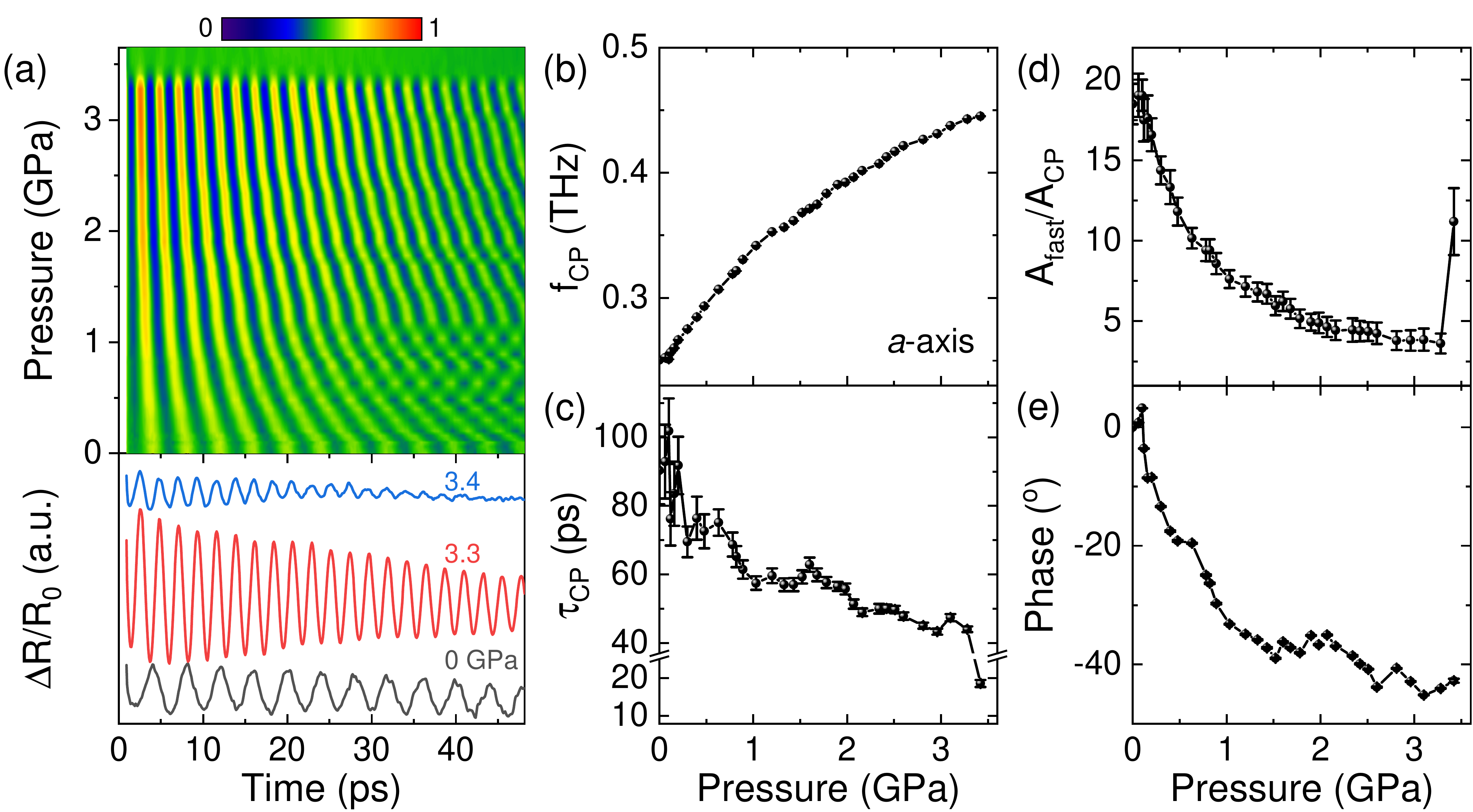}
\caption{(a) Intensity map of the coherent phonon. Time-domain intensity profiles at 0, 3.3, and 3.4 GPa are presented for clarification. (b)-(e) Pressure dependence of the frequency f$_{CP}$, lifetime $\tau$$_{CP}$, intensity ratio A$_{fast}$/A$_{CP}$, and initial phase of the coherent phonon.}
\end{figure*}

The pressure induced change in the electronic structure not only modifies the electronic response of the DR signal but also impacts the CP excitation. In Fig. 3(a), we plot the CP time-domain intensity profiles at different pressures as the color map, where the incoherent electronic background under different pressures was subtracted. Accordingly, the dynamical parameters including the frequency (f$_{CP}$), lifetime ($\tau$$_{CP}$), amplitude (A$_{CP}$), and phase ($\varphi$) versus the pressures are illustrated in the Figs. 3(b)-(e). Apparently, the CP frequency manifests blueshift with pressure, which is due to the pressure-induced enhancement of interlayer interactions. Moreover, we can see that the lifetime of CP shortens with increasing pressure, emphasizing the reinforcement of anharmonic interactions between phonons.\cite{Matsu} Such increase in an anharmonic coupling cannot fully account for the sudden vanishing of the CP oscillation above 3.5 GPa since the $\tau$$_{CP}$ should not be zero albeit it reduces by a factor of 5. Indeed, the lifetime of CP is expected to be about 40 ps at 3.5 GPa according to the trend. In addition, the CP amplitude exhibits a prominent increase before it completely disappears as shown in Fig. 3(d). The sudden changes of the lifetime and amplitude of the CP with pressures further emphasize that the topological transition of the electronic structures plays nontrivial roles for the e-ph coupling.

As shown in Fig. 3(e), the phase $\varphi$ at ambient pressure is $\sim$0, indicating a cosine-type CP oscillation, which highlights that the coherent shear phonon stems from the displacive excitation.\cite{Sie,Qi} The displacive CP excitation is associated with the shift of the potential energy surface induced by the nonradiative energy delivery through the e-ph coupling.\cite{Zeige} Thereby, we suggest that the sudden changes of CP oscillation around 3.5 GPa reflect an abrupt reduction in the e-ph coupling. It can be seen that the phase shift rapidly increases with elevating pressure and amounts to -45$^{\circ}$ around 3.5 GPa. In other word, the lattice response shows an extra delay in response to the photoexcitation under pressure. As for the displacive CP, if the lifetime of CP is far smaller than that of electronic relaxation, the phase is roughly linked with the frequency and CP lifetime, that is, $\varphi$$\approx$arctan(-1/(f$\tau$$_{CP}$ )).\cite{Riffe} Accordingly, the maximum change in the phase is estimated about 3.2$^{\circ}$ under pressure, which is much less than the experimental result. Previous work has revealed that the electron temperature and the carrier cooling during the nascent stage of electronic relaxation strongly impact the coordinate position of the potential energy surface of the excited state and can give rise to the large phase shift of CP in TiO$_2$.\cite{Boths} Along this line, we suggest that an extraordinary large phase shift has a close correlation with the changes in the electronic structure. As shown by the above calculation (Fig. 2), both the DOS and the size of the Fermi pockets increase remarkably with pressure, allowing for more state occupations at the lower energy side. As a result, the initial carrier temperature upon photoexcitation is expected to be decreased along with the ETT under pressure. Such attenuation in electron temperature may lead to a non-instantaneous driving force for CP excitation.\cite{Boths} Furthermore, we propose that the reduction both in the electron temperature and e-ph coupling due to the ETT and the formation of the Weyl point above 4 GPa decrease the energy exchange rate between electron and lattice, and further suppress the excitation of CP.

In summary, we employ the fs-OPPS in combination with DAC to investigate the pressure evolutions of nonequilibrium photocarrier dynamics in WTe$_2$, reflecting the details of the electronic structure over large k-space previously inaccessible under pressure. We find that the DR curves comprised with incoherent electronic response and CP manifest the anomalies around 0.8, 3.5, and 6 GPa, depending on the light polarization. With increasing pressure, the CP stemming from the A$_1$ mode starts to disappear above 3.5 GPa. Concomitantly, the fast time constant of the electronic relaxation shows a recognizable sudden drop around 0.8 GPa, and also a remarkable jump at around 3.5 GPa. Such pressure behaviors of photocarrier dynamics signify the electronic structure transitions in the crystal structure. The first-principles calculation reveals a topology change of Fermi surface and a new type-II Weyl energy dispersion below 6 GPa. Our work does not only develop the better understandings of the e-ph interactions in WTe$_2$, but also proposes a new technique to probe the electronic topological changes under pressure.

\begin{acknowledgments}
This work was supported by National Natural Science Foundation of China (Grant Nos. 12004387, 12174398, 11974354, 51727806), CAS Innovation Grant (No. CXJJ-19-B08) and the HFIPS Director's Fund (Grant No. YZJJKX202202).
\end{acknowledgments}


\begin{references}
\bibitem{Bansz}L. Banszerus, M. Schmitz, S. Engels, J. Dauber, M. Oellers, F. Haupt, K. Watanabe, T. Taniguchi, B. Beschoten, and C. Stampfer, Ultrahigh-mobility graphene devices from chemical vapor deposition on reusable copper, Sci. Adv. \textbf{1}, e1500222 (2015).
\bibitem{Liang}T. Liang, Q. Gibson, M. N. Ali, M. Liu, R. J. Cava, and N. P. Ong, Ultrahigh mobility and giant magnetoresistance in the Dirac semimetal Cd$_3$As$_2$, Nat. Mater. \textbf{14}, 280 (2015).
\bibitem{Ali}M. N. Ali, J. Xiong, S. Flynn, J. Tao, Q. D. Gib- son, L. M. Schoop, T. Liang, N. Haldolaarachchige, M. Hirschberger, N. P. Ong, and R. J. Cava, Large, non-saturating magnetoresistance in WTe$_2$, Nature \textbf{514}, 205 (2014).
\bibitem{Weyl}H. Weyl, Elektron und gravitation, I, Zeitschrift f\"{u}r Physik \textbf{56}, 330 (1929).
\bibitem{Yan}B. Yan and C. Felser, Topological materials: Weyl semimetals, Annu. Rev. Conden. Ma. P. \textbf{8}, 337 (2017).
\bibitem{Huang}X. Huang, L. Zhao, Y. Long, P. Wang, D. Chen, Z. Yang, H. Liang, M. Xue, H. Weng, Z. Fang, X. Dai, and G. Chen, Observation of the chiral-anomaly-induced negative magnetoresistance in 3D Weyl semimetal TaAs, Phys. Rev. X \textbf{5}, 031023 (2015).
\bibitem{Yao}M. Yao, K. Manna, Q. Yang, A. Fedorov, V. Voroshnin, B. V. Schwarze, J. Hornung, S. Chattopadhyay, Z. Sun, S. N. Guin, J. Wosnitza, H. Borrmann, C. Shekhar, N. Kumar, J. Fink, Y. Sun, and C. Felser, Observation of giant spin-split Fermi-arc with maximal Chern number in the chiral topological semimetal PtGa, Nat. Commun. \textbf{11}, 2033 (2020).
\bibitem{Xu}S.-Y. Xu, I. Belopolski, N. Alidoust, M. Neupane, G. Bian, C. Zhang, R. Sankar, G. Chang, Z. Yuan, C.-C. Lee, S. M. Huang, H. Zheng, J. Ma, D. S. Sanchez, B. K. Wang, A. Bansil, F. Chou, P. P. Shibayev, H. Lin, S. Jia,and M. Z. Hasan, Discovery of a Weyl fermion semimetal and topological Fermi arcs, Science \textbf{349}, 613 (2015).
\bibitem{Qian}X. Qian, J. Liu, L. Fu, and J. Li, Quantum spin Hall effect in two-dimensional transition metal dichalcogenides, Science \textbf{346}, 1344 (2014).
\bibitem{Jiang-1}J. Jiang, F. Tang, X. C. Pan, H. M. Liu, X. H. Niu, Y. X. Wang, D. F. Xu, H. F. Yang, B. P. Xie, F. Q. Song, P. Dudin, T. K. Kim, M. Hoesch, P. Kumar Das, I. Vobornik, X. G. Wan, and D. L. Feng, Signature of strong spin-orbital coupling in the large nonsaturating magnetoresistance material WTe$_2$, Phys. Rev. Lett. \textbf{115}, 166601 (2015).
\bibitem{Tokur}Y. Tokura, M. Kawasaki, and N. Nagaosa, Emergent functions of quantum materials, Nat. Phys. \textbf{13}, 1056 (2017).
\bibitem{Cheng}E. Cheng, W. Xia, X. Shi, Z. Yu, L. Wang, L. Yan, D. C. Peets, C. Zhu, H. Su, Y. Zhang, D. Dai, X. Wang, Z. Zou, N. Yu, X. Kou, W. Yang, W. Zhao, Y. Guo, and S. Li, Pressure-induced superconductivity and topological phase transitions in the topological nodal-line semimetal SrAs$_3$, npj Quantum Mater. \textbf{5}, 1 (2020).
\bibitem{Xiao}R. C. Xiao, P. L. Gong, Q. S. Wu, W. J. Lu, M. J. Wei, J. Y. Li, H. Y. Lv, X. Luo, P. Tong, X. B. Zhu, and Y. P. Sun, Manipulation of type-I and type-II Dirac points in PdTe$_2$ superconductor by external pressure, Phys. Rev. B \textbf{96}, 075101 (2017).
\bibitem{Xiang}Z. J. Xiang, G. J. Ye, C. Shang, B. Lei, N. Z. Wang, K. S. Yang, D. Y. Liu, F. B. Meng, X. G. Luo, L. J. Zou, Z. Sun, Y. Zhang, and X. H. Chen, Pressure-induced electronic transition in black phosphorus, Phys. Rev. Lett. \textbf{115}, 186403 (2015).
\bibitem{Jiang-2}K. Jiang, A. Cui, S. Shao, J. Feng, H. Dong, B. Chen, Y. Wang, Z. Hu, and J. Chu, New pressure stabilization structure in two-dimensional PtSe$_2$, J. Phys. Chem. Lett. \textbf{11}, 7342 (2020).
\bibitem{Zhang-2}J. L. Zhang, C. Y. Guo, X. D. Zhu, L. Ma, G. L. Zheng, Y. Q. Wang, L. Pi, Y. Chen, H. Q. Yuan, and M. L. Tian, Disruption of the accidental Dirac semimetal state in ZrTe$_5$ under hydrostatic pressure, Phys. Rev. Lett. \textbf{118}, 206601 (2017).
\bibitem{Zhang-3}C. Zhang, A. Narayan, S. Lu, J. Zhang, H. Zhang, Z. Ni, X. Yuan, Y. Liu, J.-H. Park, E. Zhang, W. Wang, S. Liu, L. Cheng, L. Pi, Z. Sheng, S. Sanvito, and F. Xiu, Evolution of Weyl orbit and quantum Hall effect in Dirac semimetal Cd$_3$As$_2$, Nat. Commun. \textbf{8}, 1 (2017).
\bibitem{Gierz}I. Gierz, J. C. Petersen, M. Mitrano, C. Cacho, I. Turcu, E. Springate, A. Stöhr, A. K\"{o}hler, U. Starke, and A. Cavalleri, Snapshots of non-equilibrium Dirac carrier distributions in graphene, Nat. Mater. \textbf{12}, 1119 (2013).
\bibitem{Othon}A. Othonos, Probing ultrafast carrier and phonon dynamics in semiconductors, J. Appl. Phys. \textbf{83}, 1789 (1998).
\bibitem{Blanc}F. Blanchard, D. Golde, F. H. Su, L. Razzari, G. Sharma, R. Morandotti, T. Ozaki, M. Reid, M. Kira, S. W. Koch, and F. A. Hegmann, Effective mass anisotropy of hot electrons in nonparabolic conduction bands of n-doped InGaAs films using ultrafast terahertz pump-probe techniques, Phys. Rev. Lett. \textbf{107}, 107401 (2011).
\bibitem{Zhang-1}K. Zhang, H. Jiang, J. Yang, J. Zhang, Z. Zeng, X. Chen, and F. Su, Pressure effects on the lattice vibrations and ultrafast photocarrier dynamics in 2\emph{H}-TaS$_2$, Appl. Phys. Lett. \textbf{117}, 101105 (2020).
\bibitem{Zhang-4}K. Zhang, J. Xie, J. Yang, H. Jiang, S. Zhang, Z. Zeng, X. Chen, T. Wang, and F. Su, Nonequilibrium electron and lattice dynamics of Sb$_2$Te$_3$ under pressure, Phys. Rev. B \textbf{105}, 195109 (2022).
\bibitem{Li}P. Li, Y. Wen, X. He, Q. Zhang, C. Xia, Z.-M. Yu, S. A. Yang, Z. Zhu, H. N. Alshareef, and X.-X. Zhang, Evidence for topological type-II Weyl semimetal WTe$_2$, Nat. Commun. \textbf{8}, 2150 (2017).
\bibitem{Soluy}A. A. Soluyanov, D. Gresch, Z. Wang, Q. Wu, M. Troyer, X. Dai, and B. A. Bernevig, Type-II Weyl semimetals, Nature \textbf{527}, 495 (2015).
\bibitem{Xia}J. Xia, D.-F. Li, J.-D. Zhou, P. Yu, J.-H. Lin, J.-L. Kuo, H.-B. Li, Z. Liu, J.-X. Yan, and Z.-X. Shen, Pressure-induced phase transition in Weyl semimetallic WTe$_2$, Small \textbf{13}, 1701887 (2017).
\bibitem{Zhou}Y. Zhou, X. Chen, N. Li, R. Zhang, X. Wang, C. An, Y. Zhou, X. Pan, F. Song, B. Wang, W. Yang, Z. Yang, and Y. Zhang, Pressure-induced Td to 1T' structural phase transition in WTe$_2$, AIP Adv. \textbf{6}, 075008 (2016).
\bibitem{Jana}M. K. Jana, A. Singh, D. J. Late, C. R. Rajamathi, K. Biswas, C. Felser, U. V. Waghmare, and C. N. R. Rao, A combined experimental and theoretical study of the structural, electronic and vibrational properties of bulk and few-layer Td-WTe$_2$, J. Phys.: Condens. Mat. \textbf{27}, 285401 (2015).
\bibitem{Song}Q. Song, H. Wang, X. Xu, X. Pan, Y. Wang, F. Song, X. Wan, and L. Dai, The polarization-dependent anisotropic Raman response of few-layer and bulk WTe$_2$ under different excitation wavelengths, RSC Adv. \textbf{6}, 103830 (2016).
\bibitem{Frenz}A. J. Frenzel, C. C. Homes, Q. D. Gibson, Y. M. Shao, K. W. Post, A. Charnukha, R. J. Cava, and D. N. Basov, Anisotropic electrodynamics of type-II Weyl semimetal candidate WTe$_2$, Phys. Rev. B \textbf{95}, 245140 (2017).
\bibitem{Jha}R. Jha, S. Onishi, R. Higashinaka, T. D. Matsuda, R. A. Ribeiro, and Y. Aoki, Anisotropy in the electronic transport properties of Weyl semimetal WTe$_2$ single crystals, AIP Adv. \textbf{8}, 101332 (2018).
\bibitem{Chen}M. Chen, K. Lee, J. Li, L. Cheng, Q. Wang, K. Cai, E. E. M. Chia, H. Chang, and H. Yang, Anisotropic picosecond spin-photocurrent from Weyl semimetal WTe$_2$, ACS nano \textbf{14}, 3539 (2020).
\bibitem{Chen-1}Y. Chen, Z. Chen, Y. Chen, L. Chen, J. Ning, B. Liu, C. Zhang, X. Lu, X. Ruan, W. Liu, P. Wang, F. Song, C. Zhang, F. Wang, J. Wu, L. He, X. Wang, R. Zhang, and Y. Xu, Anisotropic ultrafast spin/valley dynamics in WTe$_2$ films, arXiv preprint arXiv:2008.08785 (2020).
\bibitem{Wei}Y. Wei, C. Deng, X. Zheng, Y. Chen, X. Zhang, W. Luo, Y. Zhang, G. Peng, J. Liu, H. Huang, W. Cai, Q. Ge, R. Zhang, X. Zhang, and S. Qin, Anisotropic in-plane thermal conductivity for multi-layer WTe$_2$, Nano Res. \textbf{15}, 401 (2022).
\bibitem{Lu}P. Lu, J.-S. Kim, J. Yang, H. Gao, J. Wu, D. Shao, B. Li, D. Zhou, J. Sun, D. Akinwande, D. Xing, and J. F. Lin, Origin of superconductivity in the Weyl semimetal WTe$_2$ under pressure, Phys. Rev. B \textbf{94}, 224512 (2016).
\bibitem{Pan}X.-C. Pan, X. Chen, H. Liu, Y. Feng, Z. Wei, Y. Zhou, Z. Chi, L. Pi, F. Yen, F. Song, X. Wan, Z. Yang, B. Wang, G. Wang, and Y. Zhang, Pressure-driven dome-shaped superconductivity and electronic structural evolution in tungsten ditelluride, Nat. Commun. \textbf{6}, 7805 (2015).
\bibitem{Kang}D. Kang, Y. Zhou, W. Yi, C. Yang, J. Guo, Y. Shi, S. Zhang, Z. Wang, C. Zhang, S. Jiang, A. Li, K. Yang, Q. Wu, G. Zhang, L. Sun, and Z. Zhao, Superconductivity emerging from a suppressed large magnetoresistant state in tungsten ditelluride, Nat. Commun. \textbf{6}, 7804(2015).
\bibitem{Li-1}Y. Li, J. Liu, P. Zhang, J. Zhang, N. Xiao, L. Yu, and P. Niu, Electrical transport properties of Weyl semimetal WTe$_2$ under high pressure, J. Mater. Sci. \textbf{55}, 14873 (2020).
\bibitem{Cai}P. L. Cai, J. Hu, L. P. He, J. Pan, X. C. Hong, Z. Zhang, J. Zhang, J. Wei, Z. Q. Mao, and S. Y. Li, Drastic pressure effect on the extremely large magnetoresistance in WTe$_2$: Quantum oscillation study, Phys. Rev. Lett. \textbf{115}, 057202 (2015).
\bibitem{Krott}M. Krottenm\"{u}ller, J. Ebad-Allah, V. S\"{u}ss, C. Felser, and C. A. Kuntscher, Optical conductivity of the type-II Weyl semimetal WTe$_2$ under pressure, Phys. Rev. B \textbf{102}, 075122 (2020).
\bibitem{Wang}Y. Wang, K. Wang, J. Reutt-Robey, J. Paglione, and M. S. Fuhrer, Breakdown of compensation and persistence of nonsaturating magnetoresistance in gated WTe$_{2}$ thin flakes, Phys. Rev. B \textbf{93}, 121108 (2016).
\bibitem{Zhu}Z. Zhu, X. Lin, J. Liu, B. Fauqu\'{e}, Q. Tao, C. Yang, Y. Shi, and K. Behnia, Thermoelectric coefficients, and the Fermi surface of semimetallic WTe$_{2}$, Phys. Rev. Lett. \textbf{114}, 176601 (2015).
\bibitem{Ali-1}M. N. Ali, L. Schoop, J. Xiong, S. Flynn, Q. Gibson, M. Hirschberger, N. P. Ong, and R. J. Cava, Correlation of crystal quality and extreme magnetoresistance of WTe$_{2}$, Europhys. Lett. \textbf{110}, 67002 (2015).
\bibitem{Dai}Y. M. Dai, J. Bowlan, H. Li, H. Miao, S. F. Wu, W. D. Kong, P. Richard, Y. G. Shi, S. A. Trugman, J.-X. Zhu, H. Ding, A. J. Taylor, D. A. Yarotski, and R. P. Prasankumar, Ultrafast carrier dynamics in the large magnetoresistance material WTe$_2$, Phys. Rev. B \textbf{92}, 161104 (2015).
\bibitem{He}B. He, C. Zhang, W. Zhu, Y. Li, S. Liu, X. Zhu, X. Wu, X. Wang, H.-h. Wen, and M. Xiao, Coherent optical phonon oscillation and possible electronic softening in WTe$_2$ crystals, Sci. Rep. \textbf{6}, 30487 (2016).
\bibitem{Homes}C. C. Homes, M. N. Ali, and R. J. Cava, Optical properties of the perfectly compensated semimetal WTe$_2$, Phys. Rev. B \textbf{92}, 161109 (2015).
\bibitem{Ge}S. Ge, C. Li, Z. Zhang, C. Zhang, Y. Zhang, J. Qiu, Q. Wang, J. Liu, S. Jia, J. Feng, and D. Sun, Dynamical evolution of anisotropic response in black phosphorus under ultrafast photoexcitation, Nano Lett. \textbf{15}, 4650 (2015).
\bibitem{Zhang}M. Y. Zhang, Z. X. Wang, Y. N. Li, L. Y. Shi, D. Wu, T. Lin, S. J. Zhang, Y. Q. Liu, Q. M. Liu, J. Wang, T. Dong, and N. L. Wang, Light-induced subpicosecond lattice symmetry switch in MoTe$_2$, Phys. Rev. X \textbf{9}, 021036 (2019).
\bibitem{Allen}P. B. Allen, Theory of thermal relaxation of electrons in metals, Phys. Rev. Lett. \textbf{59}, 1460 (1987).
\bibitem{Malic}E. Malic, T. Winzer, and A. Knorr, Efficient orientational carrier relaxation in optically excited graphene, Appl. Phys. Lett. \textbf{101}, 213110 (2012).
\bibitem{Si}C. Si, Z. Liu, W. Duan, and F. Liu, First-principles calculations on the effect of doping and biaxial tensile strain on electron-phonon coupling in graphene, Phys. Rev. Lett. \textbf{111}, 196802 (2013).
\bibitem{Matsu}I. Matsubara, S. Ebihara, T. Mishina, J. Nakahara, N. Matsumoto, and S. Nagata, Two coherent phonon modes at the metal-insulator transition of the spinel CuIr$_2$S$_4$ compound using femtosecond pump-probe spectroscopy, Phys. Rev. B \textbf{79}, 054110 (2009).
\bibitem{Sie}E. J. Sie, C. M. Nyby, C. D. Pemmaraju, S. J. Park, X. Shen, J. Yang, M. C. Hoffmann, B. K. Ofori-Okai, R. Li, A. H. Reid, S. Weathersby, E. Mannebach, N. Finney, D. Rhodes, D. Chenet, A. Antony, L. Balicas, J. Hone, T. P. Devereaux, T. F. Heinz, X. Wang, and A. M. Lindenberg, An ultrafast symmetry switch in a Weyl semimetal, Nature \textbf{565}, 61 (2019).
\bibitem{Qi}Y. Qi, M. Guan, D. Zahn, T. Vasileiadis, H. Seiler, Y. W. Windsor, H. Zhao, S. Meng, and R. Ernstorfer, Photoinduced concurrent intralayer and interlayer structural transitions and associated topological transitions in MTe$_2$ (M= Mo, W), arXiv preprint arXiv:2105.14175 (2021).
\bibitem{Zeige}H. J. Zeiger, J. Vidal, T. K. Cheng, E. P. Ippen, G. Dresselhaus, and M. S. Dresselhaus, Theory for displacive excitation of coherent phonons, Phys. Rev. B \textbf{45}, 768 (1992).
\bibitem{Riffe}D. M. Riffe and A. J. Sabbah, Coherent excitation of the optic phonon in Si: Transiently stimulated Raman scattering with a finite-lifetime electronic excitation, Phys. Rev. B \textbf{76}, 085207 (2007).
\bibitem{Boths}E. M. Bothschafter, A. Paarmann, E. S. Zijlstra, N. Karpowicz, M. E. Garcia, R. Kienberger, and R. Ernstorfer, Ultrafast evolution of the excited-state potential energy surface of TiO$_2$ single crystals induced by carrier cooling, Phys. Rev. Lett. \textbf{110}, 067402 (2013).
\end{references}

\end{document}


\title{Probing the electronic topological transitions of WTe$_2$ under pressure using ultrafast spectroscopy}

\author{Kai Zhang}
\affiliation{Key Laboratory of Materials Physics, Institute of Solid State Physics, HFIPS, Chinese Academy of Sciences, Hefei 230031, China}
\affiliation{GBA Branch of Aerospace Information Research Institute, Chinese Academy of Sciences, Guangzhou 510700, China}
\author{Fuhai Su}\email{fhsu@issp.ac.cn}
\affiliation{Key Laboratory of Materials Physics, Institute of Solid State Physics, HFIPS, Chinese Academy of Sciences, Hefei 230031, China}
\author{Dayong Liu}\email{dyliu@ntu.edu.cn}
\affiliation{Department of Physics, School of Sciences, Nantong University, Nantong 226019, China}
\author{Wenjun Wang}
\affiliation{Anhui Key Laboratory of Condensed Matter Physics at Extreme Conditions, High Magnetic Field Laboratory, Hefei Institutes of Physical Sciences, Chinese Academy of Sciences, Hefei, 230031, Anhui, China}
\author{Yongsheng Zhang}
\affiliation{Key Laboratory of Materials Physics, Institute of Solid State Physics, HFIPS, Chinese Academy of Sciences, Hefei 230031, China}
\author{Zhi Zeng}
\affiliation{Key Laboratory of Materials Physics, Institute of Solid State Physics, HFIPS, Chinese Academy of Sciences, Hefei 230031, China}
\author{Zhe Qu}\email{zhequ@hmfl.ac.cn}
\affiliation{Anhui Key Laboratory of Condensed Matter Physics at Extreme Conditions, High Magnetic Field Laboratory, Hefei Institutes of Physical Sciences, Chinese Academy of Sciences, Hefei, 230031, Anhui, China}
\author{Alexander F. Goncharov}\email{agoncharov@carnegiescience.edu}
\affiliation{Earth and Planets Laboratory, Carnegie Institution for Science, Washington, D.C. 20015, USA}

\date{\today}
\maketitle

\section{Optical Anisotropy determined via Raman scattering}
For the Raman scattering measurements, a 660-nm sapphire laser was applied with the average power less than 2 mW measured before a $\times$20 objective lens. A single-stage spectrograph with a CCD detector (Princeton Instruments) was employed using an 1800 g/mm grating to collect the signal in the backscattering geometry.

\begin{figure*}[htbp]
\renewcommand{\thefigure}{S1}
\includegraphics[width=0.6\textwidth]{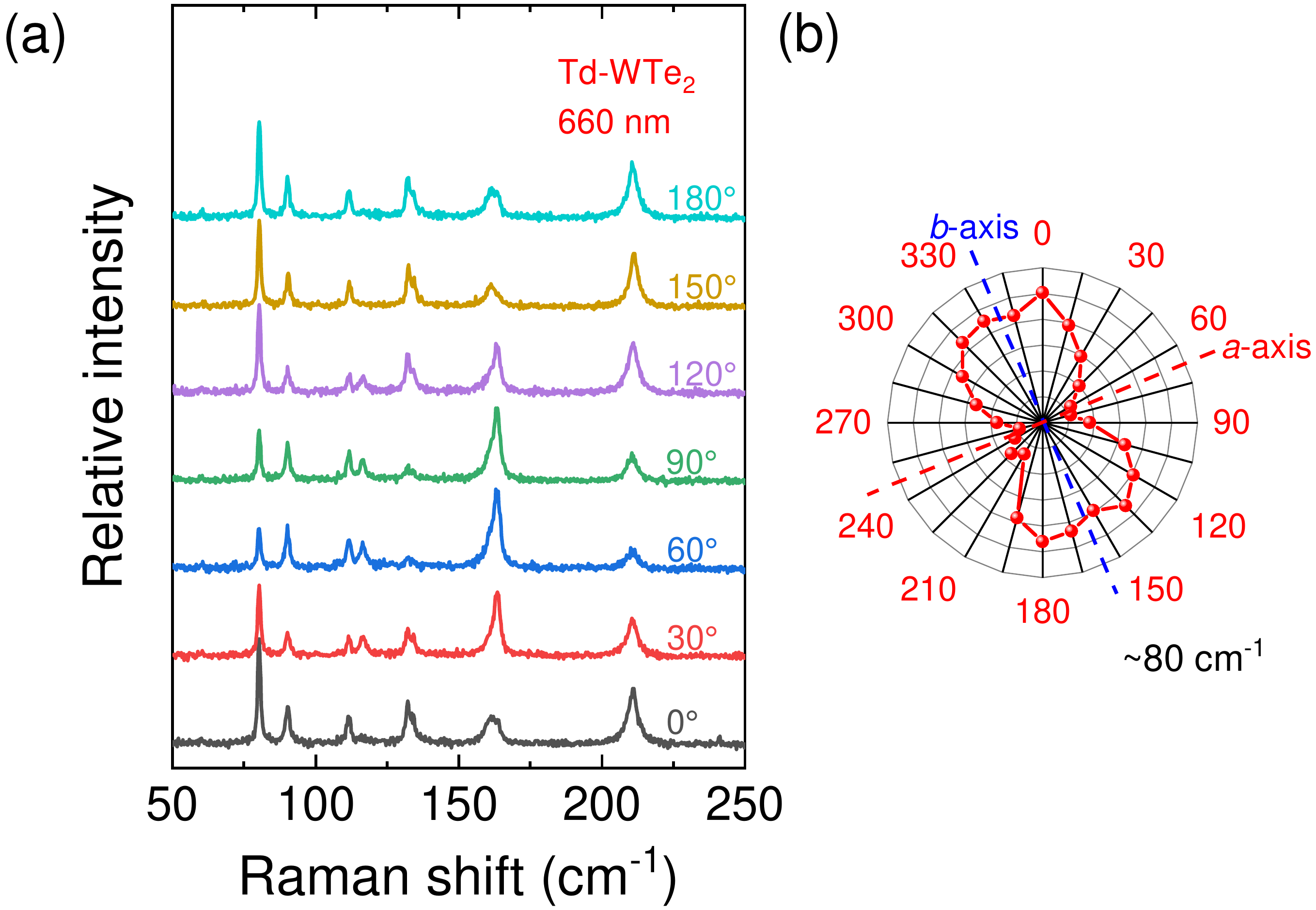}
\caption{(a) Raman spectra of Td-WTe$_{2}$ at different laser polarizations excited by a 660 nm laser. (b) Intensity plots with respect to rotation angle for the Raman mode at around 80 cm$^{-1}$. The a and b axes of the crystal lattice (dashed lines) were deduced based on these data.}
\end{figure*}

The ambient orthorhombic Td phase of the WTe$_{2}$ belongs to the C$_{2v}^{7}$ space group symmetry, the Raman active modes are 11A$_{1}$ + 6A$_{2}$.\cite{Ma,Jiang} Here, we measure the Raman spectra of the WTe$_{2}$ at different polarization directions in the ab plane to examine the anisotropy, results are shown in Fig. S1(a). Only 11 Raman modes can be observed from the spectra and each of them has been marked.\cite{Ma} Intensy of the mode at around 80 cm$^{-1}$ is extracted by fitting with Lorentz function and plotted in Fig. S1(b) with the rotation angles. Indeed, this mode shows typical twofold symmetry, which is in accordance with the previous work.\cite{Ma} The a and b axes of the host-crystal lattice can also be determined and indicated by dashed lines, here a axis is along the zig-zag tungsten chain and b axis is in the perpendicular direction, details are also painted in Fig. 1(b).

\section{Vibrational properties under pressure}

\begin{figure*}[htbp]
\renewcommand{\thefigure}{S2}
\includegraphics[width=0.9\textwidth]{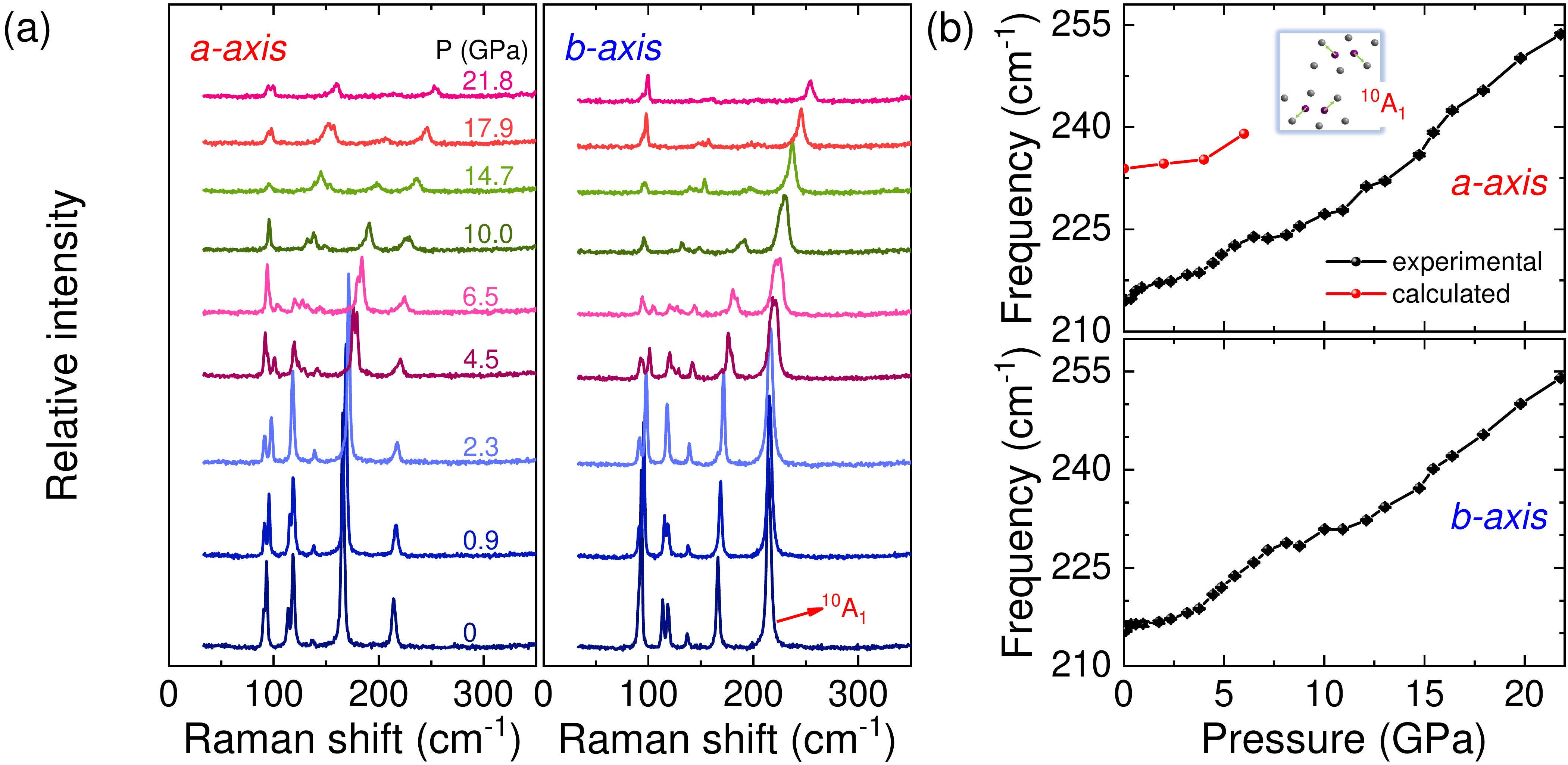}
\caption{(a) Raman spectra of WTe$_{2}$ with the laser polarization along the a and b axes of the sample at high pressures. The spectra have been displaced vertically for clarity. (b) Pressure dependence of the frequency of the $^{10}$A$_{1}$ mode obtained by the experiment (solid black lines) and theoretical calculation (solid red line).}
\end{figure*}

In order to investigate the vibrational properties of the WTe$_{2}$ at high pressures, we measure the Raman spectra with the polarization of the incident light along the a and b axes and the pressure ranging from 0 to 21.8 GPa. The spectra, as well as the frequency of the $^{10}$A$_{1}$ peak fitted by a Lorentz function, at different pressures and beam polarizations are shown in Fig. S2. The spectra along these two directions almost share the same behaviors with pressure. No obvious sudden changes in Raman shifts are observed below 20 GPa. However, the slope of the pressure dependence of the $^{10}$A$_{1}$ mode shows the subtle changes around 3.5, 7, and 15.5 GPa.

\section{Anisotropic ultrafast dynamics of the WTe$_2$}

\begin{figure*}[htbp]
\renewcommand{\thefigure}{S3}
\includegraphics[width=0.8\textwidth]{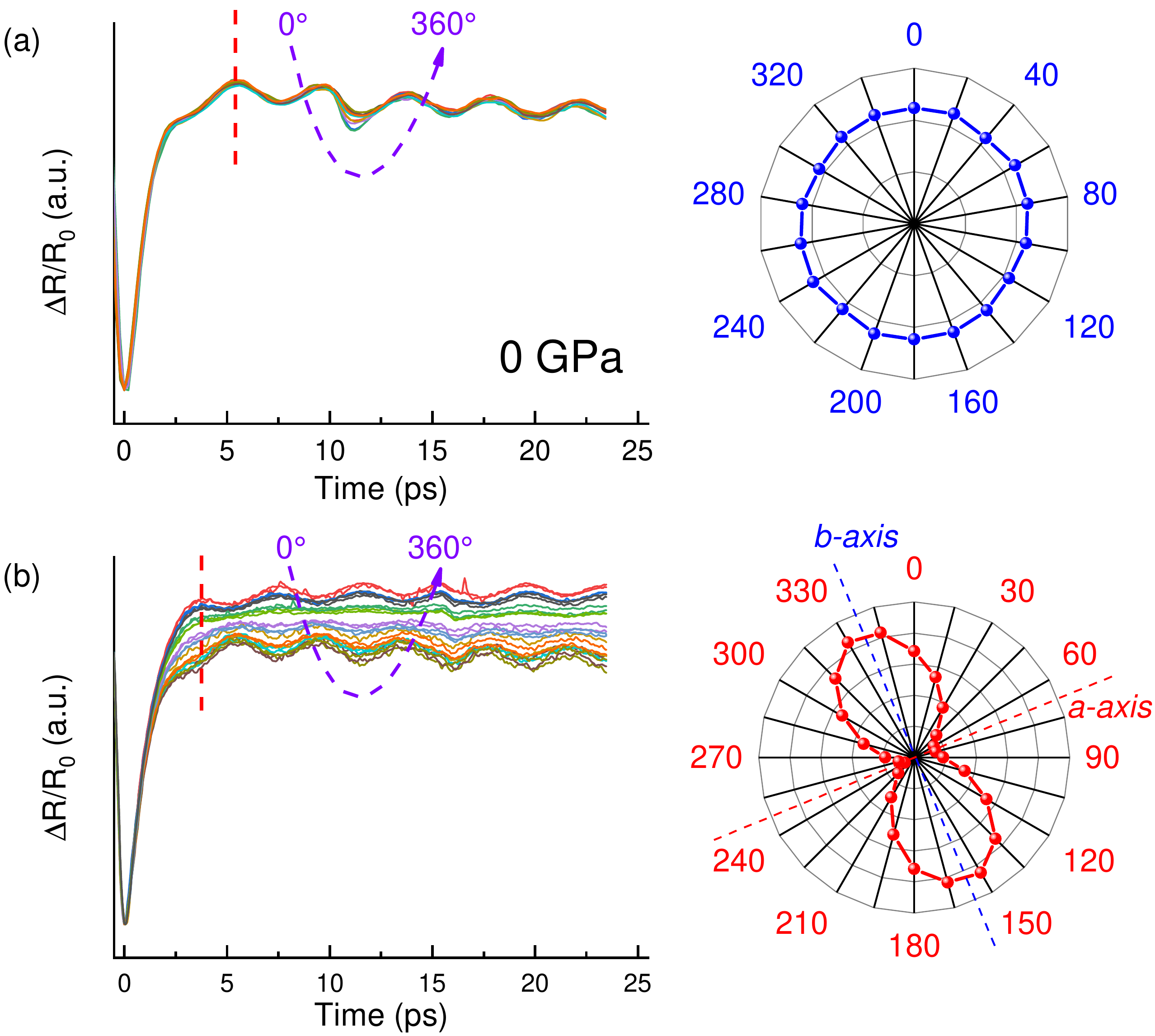}
\caption{Left panel is the normalized $\Delta$R/R$_{0}$ signal at different polarizations for the pump (a) and probe (b) beams. Right panel is the corresponding radar graphs for the angle dependent signal value at the time position indicated by vertical dashed lines in the left figures.}
\end{figure*}

Transient differential reflectivity $\Delta$R/R$_{0}$ of WTe$_2$ at different polarization of the incident pump and probe beams is presented in Figs. S3(a) and (b). The transient reflectivity falls suddenly within less than 1 ps due to the rapid increase of the hot carriers stimulated by the intense fs pump pulse. Subsequently, it starts to recover to the initial state because of the excess energy release of the unstable hot carriers to the phonon. Coherent oscillations, which superpose on the relaxation curves, can also be detected, and are demonstrated to stem from the lowest energy A$_{1}$ optical mode.\cite{Dai,He} As can be seen from Fig. S3(a), the amplitude of the oscillation frequency is almost constant with changing the polarization of the incident pump beam, indicating that the decay processes of the hot carriers are insensitive to the pump polarization. However, the amplitude of the oscillation fluctuates with changing the polarization of the incident probe laser and even disappears at certain directions. We extract the signal value at the first wave peaks/valleys and plot a radar chart with respect to rotation angle in the inset of Fig. S3. Similar with our Raman data shown in Fig. S1, WTe$_2$ shows a typical twofold symmetry, suggests a strong anisotropy exists on the electron and phonon structures of the Td-WTe$_2$. The crystallographic axes are also labeled in the inset of Fig. S3 with considering the Raman results.

\section{Computational methods}
Our first-principle calculations are performed based on density functional theory (DFT) using Perdew-Burke-Ernzerhof (PBE) functional\cite{PBE} as the exchange correlation functionals and projector augmented wave (PAW)\cite{PAW} as pseudopotentials within generalized gradient approximation (GGA) as implemented in Vienna ab initio Simulation package (VASP)\cite{VASP}. The lattice parameters under pressure are obtained through the structural relaxation with the van der Waals (vdW) correction of optB88-vdW.\cite{optB88-1,optB88-2} For high accuracy results, a large plane-wave energy cutoff of 500 eV and a dense Monkhorst-Pack\cite{MP} k-point mesh are used. The total energy and ionic force convergence criteria are set to 10$^{-6}$ eV and 10$^{-3}$ eV/${\rm \AA}$, respectively. All of the calculations are cross-checked by using both the VASP\cite{VASP} and WIEN2K\cite{Wien2k} codes.

\section{Ultrafast dynamics of the WTe$_2$ under pressure}
\begin{figure*}[htbp]
\renewcommand{\thefigure}{S4}
\includegraphics[width=0.9\textwidth]{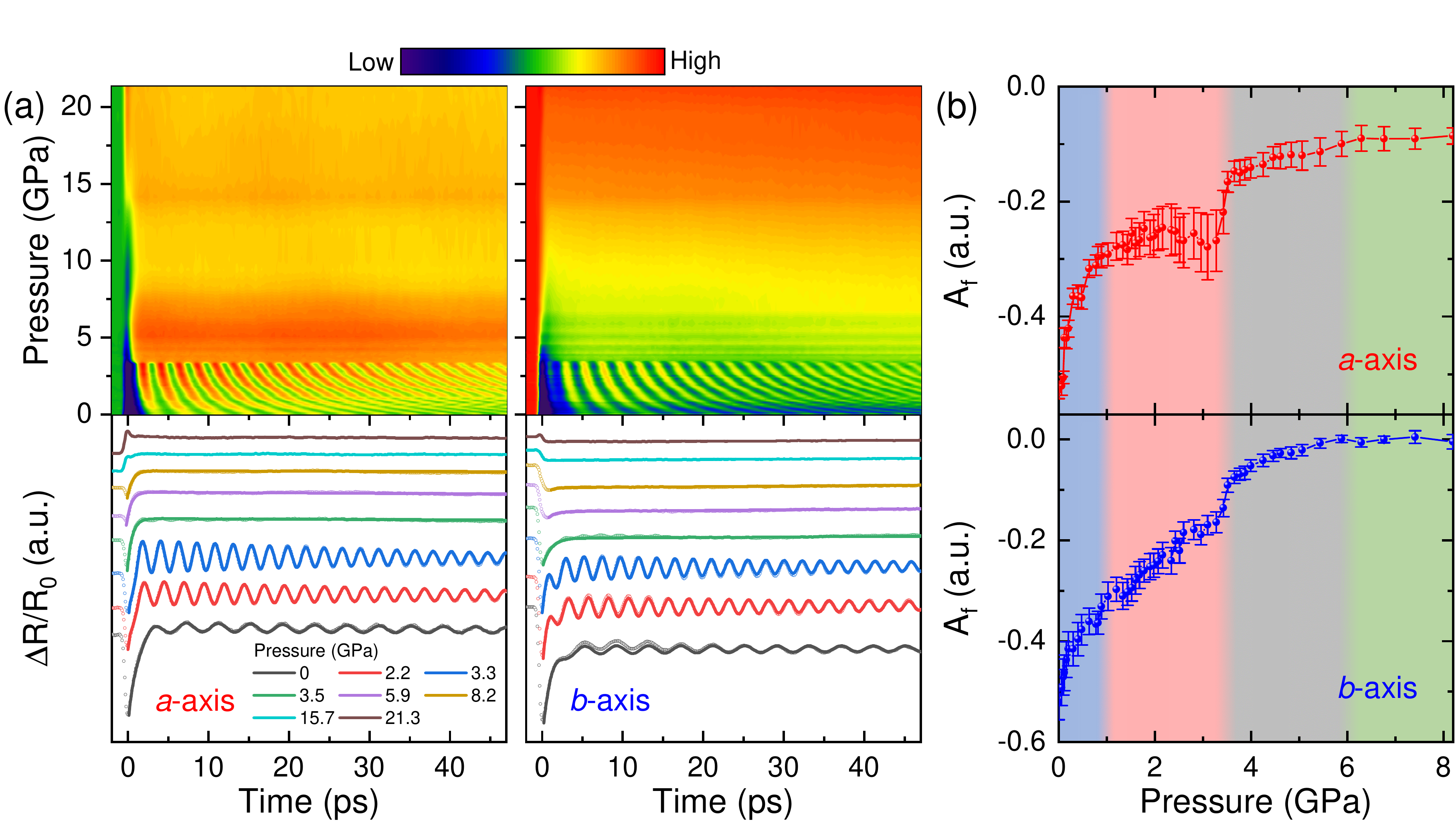}
\caption{(a) Intensity map of the transient differential reflectivity $\Delta$R/R$_{0}$ along a- and b-axes at different pressures. Raw spectra and the fitting results at several typical pressures are also presented for clarification. (b) Pressure dependence of the fast decay amplitude A$_{f}$.}
\end{figure*}
